\documentclass[12pt,a4paper]{amsart}

\usepackage{amsmath, amsfonts, xifthen, latexsym, amssymb, amsthm, amscd}

\usepackage[utf8]{inputenc}
\usepackage{graphicx}
\usepackage{url}
\usepackage{setspace}
\usepackage{epigraph}
\usepackage{hyperref}
\hypersetup{colorlinks=true,citecolor=blue,filecolor=blue,linkcolor=blue,urlcolor=blue}
\usepackage[margin=1.4in]{geometry}

\newtheorem{theorem}{Theorem}

\newtheorem{conjecture}{Conjecture}[section]

\theoremstyle{definition}
\newtheorem{definition}{Definition}[section]

\newtheorem{remark}{Remark}

\newtheorem{example}{Example}

\newcommand{\eps}{{\varepsilon}}

\renewcommand{\phi}{{\varphi}}

\newcommand{\cB}{\mathcal{B}}\newcommand{\cP}{\mathcal{P}}

\newcommand\R{\mathbb R}

\newcommand\Z{\mathbb Z}
\newcommand\F{\mathbb F}

\newcommand\cG{\mathcal{G}}

\newcommand{\actson}{\curvearrowright}

\newcommand{\cC}{\mathcal{C}}

  \newcommand{\dist}{\operatorname{dist}}

\newcommand{\vi}{\vskip 0.1in \noindent}
\usepackage{etoolbox}
\newtoggle{no_cases}\toggletrue{no_cases}
\newcommand{\case}[2][]{\iftoggle{no_cases}{\left\{\begin{array}{ll}#2 & #1}{\\#2 & #1}\togglefalse{no_cases}}
\newcommand{\esac}{\end{array}\right.\toggletrue{no_cases}}

\newcommand{\Prob}{\operatorname{Prob}}

\newcommand{\Cay}{\operatorname{Cay}}

\newcommand{\cgdk} {C\cG_{d,K}}
\begin{document}

\title[Very large graphs]{Learning Very Large Graphs with Unknown Vertex Distributions}
\author{Gábor Elek}
\address{Department of Mathematics And Statistics, Fylde College, Lancaster University, Lancaster, LA1 4YF, United Kingdom}

\email{g.elek@lancaster.ac.uk}

\begin{abstract} Recently, Goldreich introduced the notion of property testing of bounded-degree graphs
with an unknown distribution \cite{Goldreich}.  We propose a slight modification of his idea:
the Radon-Nikodym Oracles. Using these oracles any reasonable graph property can be tested in constant-time
against any reasonable unknown distribution in the category of planar graphs. We also discuss Randomized Local
Distributed Algorithms, which work on very large graphs with unknown distributions. Finally, we discuss
how can we learn graph properties using observations instead of samplings. \end{abstract}
\maketitle
\noindent
\newpage
\tableofcontents
\newpage
\begin{center} \textbf{\large Foreword} \end{center}
The following note is about Property Testing of Bounded Degree Graphs Against Unknown Vertex
Distributions, a subject recently developed by Goldreich \cite{Goldreich} (see also \cite{Goldreich2}). 
We propose a slightly different
oracle as given in the original paper. The idea goes as follows. \vi
We want to learn a very large finite graph via samplings.
However, the graph is equipped with a vertex distribution unknown for us.
 We sample the vertices according to the unknown distribution
and try to learn, say, whether the graph is far from being a forest or far from having no
$4$-cycles. Note that the statement that our graph is far from being a forest or having no $4$-cycle is relative: 
it depends on the unknown distribution. So, somehow we need to learn not only about the graph but the distribution itself.
The actual vertex probabilities are so incredibly small that sampling them is impossible. Therefore we make the following 
innocent 
assumption: there exists some constant $K>0$ such that for any adjacent vertices of our graphs the ratio of the probabilities
of the vertices is less or equal then $K$ (so, we cannot allow zero probability, but we can allow that the probabilities
are concentrated in a very small part of our graph). Our oracle will sample a vertex and explore a certain neighbourhood of the vertex. The only data about the distribution the oracle can collect is the relative values of the probabilities
in the given small neighborhood, so numbers that are not very large or very small. Basically: we sample the geometry of the bias.
It turns out that for {\bf planar graphs all reasonable graph properties can be tested in constant-time against all reasonable 
 vertex distributions}. 

\vi Note that a very similar result has been proved in the uniform case by Hassidim, Kelner, Nguyen and Onak \cite{HKNO} 
and also, Newman and Sohler \cite{Sohler}. In the uniform case, the so-called hyperfinite graph classes are the best, 
in terms of property testing and parameter estimation. In the case of the unknown distibution an interesting subfamily of the 
hyperfinite graph classes will play the same role: the Property $A$ graphs.
Finally, we discuss how can we learn graph properties for all possible vertex distribution {\it without any actual sampling} using observations.

\vi
At the end of the note there are some short sections about dynamical systems, since the property testing idea itself
leads some interesting applications in this field. 

\newpage

\section{Very  Large Finite Graphs}
{\bf What does a ``very large finite graph mean''?}
There are less than $10^{60}$ atoms in the Universe. If each and every of them could work as a supercomputer making $10^{30}$ floating point operations
per second for $10^{20}$ years, then they surely cannot input more than $10^{200}$ bits during the possible lifetime of the Universe. So, a graph of more than
$10^{200}$ vertices, let alone $10^{10^{200}}$ vertices is, in any practical sense, infinite. However, there is a hypothetical way to learn even these immensely
huge graphs using statistical samplings. If we had random access to the vertices of a very large finite graph $G$ of vertex degree bound $10$, then
by the law of large numbers, we would learn with very high probability, up to a excellent precision the percentage of vertices of degree $4$ in $G$ or
even the percentage of vertices in $G$ which are in a triangle. Well, if we had random {\it uniform} access... 
\vi
What if, we had a random access to the vertices of the graph according to an unknown distribution? Before getting into any sort of
details, let us look at a simple example.
\begin{example}
For a very large $n$, let $T_n$ be a binary tree of depth $n$.
So, we have $n$ layers of vertices $\{L_i\}^{n-1}_{i=0}$, where the layer $L_i$
contains $2^i$ vertices. For each vertex of the layer $L_i$, $2\leq i \leq n-2$,
there are two neighbours in the layer $L_{i+1}$ and one neighbour in the
layer $L_{i-1}$. Let $\beta\geq 0$ be a real number and
suppose that all the vertices $x$ in the layer $L_k$ have weight 
$w^\beta(x)=\exp(-\beta k)$. Then, we have a probability distribution 
$p^\beta:V(T_n)\to \R^+$ on the vertices, where
$$ p^\beta(x)=\frac{w^\beta (x)}{\sum_{y\in V(T_n)} w^\beta(y)}\,.$$
\end{example}
\noindent
Observe that $p^0$ is the uniform distribution. If we pick
a vertex $x$ according to $p^0$, the probability that our vertex
is of degree $1$ is basically $\frac{1}{2}$. The probability to pick
a vertex far from such vertices of degree $1$ is exponentially small.
If $\beta$ is increasing, then the probability
to pick a vertex of degree $1$ is decreasing. \vi At the critical value
$\beta=\log(2)$, the probability to pick a vertex of degree $1$ is merely
$\frac{1}{n}$, in fact, for any $0\leq i\leq n-1$, the probability
to land in the layer $L_i$ is the same. \vi
If we are passing through the critical value, that is, $\beta>\log(2)$, 
the probability starts to concentrate around the top layer. That is,
for any $\epsilon>0$, there exists some constant $K_{\epsilon,\beta}>0$
such that the probability to land in $\cup_{i=0}^{K_{\epsilon,\beta}} L_i$ is larger
than $1-\epsilon$. If $\beta$ approaches infinity the probability
is more and more concentrating onto the top vertex.
So, under the critical value we feel the ``boundary'' of our tree, above the critical value
we feel the root and at the critical value $\beta=\log(2)$ we have a {\bf phase transition},
we sort of feel an infinite $3$-regular tree. The change of the probability is changing our
picture about our sampled graph. In other words, the constant $\beta$ provides us Alternate Realities
for the same trees. In some Alternate Realities one can see the ``boundary'' in some Alternate Realities
one can see concentration around the root, and in one specific Alternate Reality one sees a homogeneous structure.
\section{The Sampling Process with Uniform Access} \label{sec2}
Suppose that we have a very large graph $G$ with some small vertex degree bound $d$, equipped with the uniform probability measure.
We collect information about $G$ using an oracle in our sampling process. The oracle accesses a vertex $x$ randomly and
explores its $r$-neighbourhood. What does it mean? The data the oracle collects is the ball
$B_r(G,x)$ of radius $r$ centered around $x$. There are only finitely many such balls (up to rooted isomorphism) 
in a graph of vertex degree bound $d$. Denote this set by $\cB^{d,r}$. Using repeated samplings of the graph by our oracle,
 we can estimate the percentage of the vertices $x$ in $G$ up to high precision, such that the ball of
radius $r$ centered around $x$ is isomorphic to a certain element $B\in \cB^{d,r}$. So, even knowing only
a miniscule part of the graph $G$ we can guess very well the following quantity:
$$ \Prob_G(B):=\frac{|\{x\in V(G)\mid \, B_r(G,x)\,\mbox{is isomorphic to $B$}\,\}|}{|V(G)|} $$
\noindent
The point is that for any $\epsilon>0$ and $k>0$, there is a constant $C_{d,\epsilon,k}>0$ such
that using $C_{d,\epsilon,k}$ sampling queries the probability that our empirical guess for any test ball
 $B\in \cB^{d,r}$ 
differs from the actual value  $\Prob_G(B)$ by more than $\epsilon$ is less than 
$\epsilon$, independently on the size of the graph $G$. The idea to sample graphs in this way is due to
 Goldreich and Ron \cite{GR}, based on earlier work on general graph property testing by Goldreich,
Goldwasser and Ron \cite{GGR}.
\section{The Miracle of Constant-Time Algorithms}
Let $G$ be a finite graph and $M(G)\subset E(G)$ be a maximum sized matching in $G$.
The {\bf matching number} is defined as
$$m(G):= \frac{|M|}{|V(G)|}.$$
\noindent
The fastest algorithm (due to Micali and Vazirani \cite{Micali}) to compute $m(G)$ has computational 
time $K_d |V(G)|^{3/2}$, where the constant
$K_d$ depends on the degree bound $d$.
Now, suppose that we do not want to compute the exact value of $m(G)$, only an approximate value
$m_{app}(G)$ satisfying the inequality
\begin{equation} \label{eq1}
|m_{app}(G)-m(G)|<\epsilon,
\end{equation}
\noindent
where $\epsilon>0$ is a very small constant, say $\eps=10^{-10000}$. Also, if a graph $G$
is given we only require our guess value $m_{app}(G)$ to satisfy \eqref{eq1} with probability at least $1-\eps$.
Then, we do not need polynomial-time, linear-time, or even logarithmic-time. We only need: constant-time.
That is, there exist constants $C(\eps,d)>0$ and $K(\eps,d)>0$ depending on $\eps$ and the degree bound $d$
(and {\it not} on the size of the graph!) such that if our oracle queries the graph $G$
of degree bound $d$\, $C(\eps,d)$-times and explore the $K(\eps,d)$-neighborhood of the sampled
vertices, then based on the date collected by the oracle, we can actually compute a constant $m_{app}(G)$
satisfying \eqref{eq1} with probability larger than $1-\eps$ (see \cite{NO}).
In other words, the matching number as a graph parameter can be tested (estimated) in constant-time.
\vi
One should note that some very important graph parameters cannot be tested in constant-time.
Let $G$ be a graph as above and let $I(G)\subset V(G)$ be a maximum sized independent set of $G$.
The independence number of $G$ is defined as
$$i(G):=\frac{|I(G)|}{|V(G)|}\,.$$
\noindent
Then, in the category of finite graphs of degree bound $3$, the independence number $i(G)$ cannot be
tested in constant-time. A short argument for this goes as follows. With high probability,  large, random $3$-regular graphs have
independence number less than $0.49$ \cite{Bollobas} and they cannot be locally distinguished from a
random $3$-regular bipartite graph (having independence number $0.5$). If we pick a vertex $y$ of any such very large graphs
the $k$-neighborhood of $y$ will be the same rooted tree of depth $k$. Therefore, knowing the local
statistics of the $k$-neighbourhoods cannot help us to estimate the parameter $i(G)$.
\section{Property Testing. The classical case}
Let $\cG^d$ be the set of all finite graphs $G$ with bounded degree $d$, up to isomorphism.
By a property $\cP$, we just mean a subset of $\cG^d$ such as forests, planar graphs, graphs without $4$-cycles.
The {\bf edit distance} of the graphs $G,H\in \cG^d$ of the same size is defined as
$$\dist_e(G,H)=\min_{H'}\frac{E(G)\triangle E(H')}{d |V(G)|}\,,$$
\noindent
where the minimum is taken for all graphs $H'$ on the vertex set of $G$ which are isomorphic to $G$.
So, the edit distance measure the percentage of edges one should change to obtain $H$ from $G$.
We can define $\dist_e(G,\cP)$ by
$$\dist_e(G,\cP):=\min_{H\in\cP,\, |V(H)|=|V(G)|} \dist_e(G,H)\,.$$\
\noindent
Following Goldreich and Ron \cite{GR} we say that the property $\cP$ is {\bf testable}, if
for any $\epsilon>0$ there exists some $r>0$ and a constant $C_{r,\eps}>0$ such that after
making $C_{r,\eps}$ queries on the vertices by an oracle exploring $r$-balls around chosen vertices we
can ACCEPT $G$ or REJECT $G$ in the following way.
\begin{itemize}
\item If $G\in \cP$ we must ACCEPT $G$.
\item If $\dist_e(G,\cP)>\eps$, we must REJECT $G$ with probability more than $1-\eps$.
\end{itemize}

\section{Hyperfinite graphs}
Let $\cG\subset \cG^d$ be a subclass of finite graphs.  We say that the class $\cG$ is
{\bf hyperfinite} (see \cite{elekhyper}) if for any $\epsilon>0$ there exists a constant $K_\epsilon>0$ such that for any
$G\in\cG$, we have a subset $Y\subset V(G)$ satisfying the following two conditions.
\begin{itemize}
\item $|Y|<\epsilon |V(G)|.$
\item If we delete $Y$ and all edges incident to a vertex in $Y$, the remaining graph $G'$ has components
of size at most $K_\epsilon$.
\end{itemize}
\noindent
Many graph classes of the ``real world'' are actually hyperfinite. For an example, the class of planar
graphs is hyperfinite. Also, if all the graphs in a class $\cG$ have polynomial growth, that is
there exists some polynomial $P$ such that for any $G$ and any $v\in V(G)$ we have $|B_r(G,v)|<P(r)$, then
the $\cG$ is hyperfinite as well. For hyperfinite graph classes, any reasonable
graph parameter and property can be testable in constant-time 
(see \cite{Sohler},\cite{HKNO},\cite{elekamena}). For the purpose of this paper ``reasonable'' means (to be on the safe side)
a graph property closed under taking subgraphs and disjoint unions, e.g. being bipartite, or planar, or a forest, or $k$-colorable.
\section{The Case of an Unknown Distribution}\label{sec6}
Now, we assume that our oracle has access to graphs $(G,p_G)$ of vertex degree bound $d$, equipped with
an unknown probability distribution $P_G:V(G)\to \R^+$. The oracle pick a vertex $x\in V(G)$ according to the
law $p_G$ and collect  the $r$-ball $B_r(G,x)$ as a result of the query. 
Goldreich \cite{Goldreich} (see also \cite{Goldreich2}) to evaluate the value $p(x)$ as
well. This is the point, where we propose a small digression from the original definition. 
In Example 1., when our parameter $\beta$ is smaller or equal than the critical value
most of the points (according to the law $p^\beta$) have incredibly small probability for large $n$ values. As
$n$ goes to infinity it is harder and harder to store the data. However, the ratio $\frac{p^\beta(x)}{p^\beta(y)}$
that is, not the {\it absolute}, but the {\it relative} probabilities in the sampled balls, stays bounded for 
adjacent vertices. So, we propose the following definition.
\begin{definition}[The Radon-Nikodym Oracle]
Let $\cG$ be a class of finite graphs $(G,p_G)$ of vertex degree bound $d$ equipped
with some probability measure $p_G$.
Assume that there exists some global constant $K>1$
such that if $G\in\cG$ and $v,w$ are adjacent vertices, then
\begin{equation} \label{boundeq}\frac{1}{K}\leq \frac{p(v)}{p(w)} \leq K\,.\end{equation}
\noindent
The Radon-Nikodym Oracle pick a vertex $x$ explore the ball $B_r(G,x)$ and the result of the query is
a labeled copy of the rooted ball $B_r(G,x)$, where the label of $y\in B_r(G,x)$ is
$l(y)=\frac{p_G(y)}{p_G(x)}$. 
\end{definition}
\noindent
Since we cannot really store a real number, we introduce a technical parameter $t$.
An oracle of depth $t$ store the the actual label $l(y)$ only up to the first $t$ digits (after the decimal point).
E.g. an oracle of depth $2$ store $3.14$ instead of the $\pi$.
It is important to observe that by our assumption, the potential result of a query of such an Radon-Nikodym Oracle of
depth $t$ is in a finite set of labeled balls $\cB^{d,r,K,t}.$ From now on we call measured finite graphs $(G,p_G)$ as
above $K$-weighted graphs. 
\begin{remark}
We can relax the boundedness condition \eqref{boundeq} by assuming that the set of vertices $x$ in the graphs $G\in \cG$
for which $\frac{1}{M} \leq \frac{p(y)}{p(x)} \leq M$ tends to zero as $M$ tends to infinity. For these {\bf tight}
graph families the Radon-Nikodym Oracles behave as well as for $K$-weighted graphs.
\end{remark} 

\section{The Discrete Radon-Nikodym Derivative}
It seems that we owe some explanation for naming our oracle a ``Radon-Nikodym Oracle''.
Let us suppose that we have a very large graph $G$, equipped with a probability measure $p_G$ satisfying
the estimate \eqref{boundeq} for some $K$.
Let us label every directed edge $e=\overrightarrow{(v,w)}$ by the number $\frac{p_G(w)}{p_G(v)}$. So, the the label $r(e)$ measures
the rate of change in the probability $p_G$ along the directed edge $e$.
Let $A,B\subset V(G)$ be subsets and $\phi:A\to B$ be a bijection such that if $a\in A$, then $a$ and $\phi(a)$
are adjacent vertices. Then, 
$$p_G(B)=\sum_{v\in A} r(v,\phi(v)) p_G(v)=\int_A r(v,\phi(v)) d\,p_G(v)\,.$$
\noindent
That is, the edge function behave like the Radon-Nikodym derivative (see Section \ref{secradon}). In fact, our
oracle sample exactly the Discrete Radon-Nikodym derivatives. 

\section{Property Testing of Graphs With An Unknown Vertex Distribution} \label{property}
Finally, we state the definition of property testing of graphs against an unknown vertex distribution.
Let $(G,p_G)$ and $(H,p_H)$ be two $K$-weighted graphs such that $|V(G)|=|V(H)|$.
If $e$ is an edge of $G$, set $wp_G(e)= p_G(u)+p_G(v)$, where $u,v$ are the endpoints of $e$.
The $K$-weighted edit distance of $G$ and $H$ is defined in the following way.
$$\dist_K((G,p_G),(H,p_H))=\sum_{e\in E(G)\backslash E(H)} wp_G(e)+ \sum_{f\in E(H)\backslash E(G)} wp_H(f)\,.$$
\noindent
Finally, the $K$-weighted edit distance of $(G,p_G)$ from the property $\cP$ is defined as 
 $$\dist_K((G,p_G),\cP)=\inf_{(H,p_H)} \dist_K((G,p_G),(H,p_H))\,,$$
\noindent
where the infinum is taken on all $K$-weighted graphs $H$ on the vertex set of $G$.
\vi
We say that {\bf a property $\cP$ is testable in constant-time with a Radon-Nikodym Oracle  in a graph class $\cG$ for any reasonable unknown 
vertex distribution} if
for any $\epsilon>0$ and $K>0$, there exist $t>0$, $r>0$ and a constant $C_{\eps,K}>0$ such that the following machinery works.
First we make an agreement with an adversary in the constants $\eps$ and $K$.  Then, our adversary gives us a graph $G\in\cG$ and chooses a vertex distribution $p_G$ on $V(G)$ at his pleasure 
that satisfies
\eqref{boundeq} for the constant $K$. 
Now, our Radon-Nikodym Oracle explores the $r$-balls around $C_{\eps,K}$ vertices
sampled according to the  probability $p_G$. Based on the findings, that is, $C_{\eps,K}$ balls from the set $\cB^{d,r,K,t}$,
the oracle ACCEPT of REJECT the graph $G$ in such a way that:
\begin{itemize}
\item The oracle accept $G$ if $G\in\cP$.
\item If $\dist_K((G,p_G),\cP)>\eps$, the oracle must REJECT $G$ with probability more than $1-\eps$.
\end{itemize}
\noindent
It is important that the assumptions should be satisfied no matter which probability our adversary chooses. Note that
it is possible that for some probability distribution $p^1_G$, $\dist_K((G,p^1_G),\cP)<\eps$ and for some probability
distribution $p^2_G$,  $\dist_K((G,p^2_G),\cP))>\eps$. The point is that the probabilities $p_G$ provide possible Alternate 
Realities for our graph $G$ (as it explained in our favourite Example 1.) and the oracle should learn the the Alternate Truth about the 
graph in a coherent
fashion according to the given Alternate Reality. Is there an Absolute Truth about very large graphs? We will discuss this question in
Section \ref{absolute}.

\section{For Planar Graphs All Reasonable Graph Properties Are Testable} \label{sec9}
A class $\cG$ of $K$-weighted finite graphs $(G,p_G)$ of degree bound $d$ is called {\bf weighted hyperfinite}
\cite{Elektimar}, if for any $\epsilon>0$ there exists $K_\epsilon>0$ such that for any
$(G,p_G)\in\cG$ one can find a subset $Y\subset V(G)$ so that
\begin{itemize}
\item $p_G(Y)\leq \eps$.
\item If we delete the subset $Y$ along with all the incident edges,
all the components of the remaining graph $G'$ have at most $K_\epsilon$ elements.
\end{itemize}
\noindent
By the result of Sako \cite{Sako}, any graph class $\cG$ satisfying the so-called Property $A$ is
weighted hyperfinite with respect to {\it any} probability measure satisfying \eqref{boundeq} for a fixed $K$.
Since planar graphs are of Property $A$ \cite{Ostrovskii},
we can conclude that $K$-weighted planar graph classes are always weighted hyperfinite (we thank
Ana Khukhro to call our attention to the paper \cite{Ostrovskii}). The same holds
for classes of polynomial (or even subexponential) growth. Using the notion of weighted hyperfiniteness and
a technique similar (but a bit more involved) as in \cite{elekamena} we can prove our main result. 
\begin{theorem}\cite{EK}
For planar graphs all reasonable graph properties against all reasonable vertex distributions are testable in constant-time.
\end{theorem}
\noindent
The result also holds for any Property A graph classes such as graphs of given polynomial or even subexponential growth.
\vi
Let $(G,p_G)$ be a finite $K$-weighted graph. Let $I\subset V(G)$ be an independent set of maximal probability.
That is, for any independent set $J\subset V(G)$, we have
$$\sum_{x\in J} p_G(x)\leq \sum_{y\in I} p_G(y)\,.$$
\noindent
Then, $i(G,p_G):=  \sum_{y\in I} p_G(y)$ is called the {\bf independence number of $G$ with respect to $p_G$}. Testability
(or estimability) of
the independence number with an unknown vertex distribution in a given graph class $\cG$ means the following. Again,
we agree in an $\epsilon$ and a $K$ with our adversary. Then, the
adversary chooses a graph $G$ from the class $\cG$ and a probability distribution on $V(G)$. 
Our oracle carries through exactly the same sampling process as before
and  produces an approximative answer $i_{app}(G,p_G)$ in such a way that with probability more than $1-\epsilon$,
$$|i(G,p_G)-i_{app}(G,p_G)|<\eps$$
\noindent
holds.
\begin{theorem}\cite{EK}
For planar graphs (or any other Property A graph class) the independence number is testable in constant-time with Radon-Nikodym Oracles.
\end{theorem}
\section{Weighted Benjamini-Schramm Convergence}
First let us recall the definition of the classical Benjamini-Schamm convergence \cite{BS} for unweighted graphs.
Let $\{G_n\}^\infty_{n=1}$ be an increasing sequence of finite graphs of vertex degree bound $d$.
We say that the sequence is {\bf convergent}, if for any $r\geq 1$ and ball $B\in \cB^{d,r}$ (see Section \ref{sec2})
the probabilities $\Prob_{G_n}(B)$ converge. E.g. the trees $\{T_n\}$ of Example 1. equipped with the uniform 
measure converge in the sense of Benjamini and Schramm.
Now, let $\{G_n,p_n\}^\infty_{n=1}$ be an increasing sequence of $K$-weighted graphs. 
Recall from Section \ref{sec6} the finite set of rooted-labeled balls $\cB^{d,r,K,t}$.
\begin{definition}[Take it with a grain of salt] \label{weight}
The sequence $\{G_n,p_n\}^\infty_{n=1}$ of $K$-weighted graphs is convergent in the sense of Benjamini and Schramm, if
for any $r\geq 1$,  $t\geq 1$ and edge-labeled ball $B\in \cB^{d,r,K,t}$,
the probabilities $\Prob_{G_n}(B)$ converge.
\end{definition}
\begin{remark}
The  precise definition of weighted Benjamini-Schramm convergence involves the notion of weak convergence 
of measures on a certain compact space and will be given in Section \ref{sec13}. The 
definition above is almost correct, save in the case, when there is a concentration for a certain rational
number having only finitely many nonzero digits in the Discrete Radon-Nikodym Derivative. Say, for any large $n$ and $\eps>0$ the total probability
of vertices $x\in V(G_n)$ for which there exists an adjacent vertex $y$ such that $2-\eps<\frac{p(y)}{p(x)}<2+\eps$
is greater than $\frac{1}{10}$. Then, it is possible that the graphs are convergent in the ``precise'' sense, but
for odd values of $n$ all the fractions above will be slightly smaller than $2$ and for even values of $n$ 
all the fractions will be slightly greater than $2$, hence Definition \ref{weight} does not detect convergence.
If there is no such concentration around a rational number, then our definition is correct. 
\end{remark}
\noindent
It is not hard to see that any sequence of $K$-weighted graphs contains a convergent graph sequence.
So, the notion of convergence is precompact and can be metrize with a distance: the statistical distance $d_S$ (see 
\cite{elekamena} for the unweighted case). The further two $K$-weighted graphs are in the statistical distance, the
easier to distinguished them using Radon-Nikodym Oracles. In fact, 
a property $\cP$ is always testable in a graph class $\cG$ if the distance from the property with all the possible
probability distributions satisfying \eqref{boundeq}
is a continuous function with respect to the statistical distance. 
\vi
\section{Randomized Local Distributed Algorithms on $K$-weighted Graphs}
One of the goal of this note is to convince computer scientists that 
Radon-Nikodym Oracle is the right approach towards graphs with an unknown distribution. This short
section is intended to make a further point. 
Randomized Local Distributed Algorithms in the uniform case works as follows. Let us suppose that
we have a graph $G\in\cG^d$ and we want not only to estimate the independence number in constant-time, but to
build a near-maximum independent set in constant-time. Each vertex $x$ obtains a ``manual'' and explores its own
$r$-neighborhood. If the ball $B_r(G,x)$ is in the manual, then $x$ decides to be a member if the near-maximum
independent set, if not, $x$ decides not to be a member of the set. In order to ``break'' the possible symmetries,
$x$ takes a bounded amount, say $k$, of random bits and, in fact, the manual contains balls vertex-labeled
with $\{0,1\}^k$. This process is called
a Randomized Local Distributed Algorithm. Hassidim, Kelner, Nguyen and Onak showed in \cite {HKNO} that for 
hyperfinite graph classes one has randomized distributed algorithms to produce a near-maximum independent set
(and many other instances as well) with high probability.
Say, we have the class of planar graphs (or any other Property $A$ class). Then, we can use the 
Radon-Nikodym Oracle of large enough depth $t$ to produce such randomized distributed algorithms, where ``randomized'' refers to the
random seeds and nothing to do with the probabilities of the vertices. Now, the manual contains elements of
 $\cB^{d,r,K,t}$ with extra vertex labels from the set $\{0,1\}^k$ and $r$ and $t$ depend on $\epsilon$ and $K$. The process is exactly the same as in the
uniform case. 
\begin{theorem} \cite{EK}  For any $\epsilon >0$ and $K>0$ we have a Randomized Local Distributed Algorithm which
for any planar graph and any probability distribution $p_G$ 
construct independent and $J\in V(G)$ such that $|i(G,p_G)-\sum_{v\in J} p_G(v)|<\epsilon$ with probability more than $1-\epsilon$.
\end{theorem}
\section{A Remark on Small Perturbations}
Let $\{G_n\}^\infty_{n=1}$ be a sequence of larger and larger paths. This is, arguably, the simplest example of a
hyperfinite family. Clearly, all reasonable graph properties can be tested against all reasonable distributions on this family. 
Now, suppose that $|V(G_n)|=n^2$. Let $\{H_n\}^\infty_{n=1}$ be an expander sequence, where $|V(H_n)|=n$.
Let us construct a graph sequence $\{J_n\}^\infty_{n=1}$ in the following way. 
\begin{itemize}
\item $V(J_n)=V(G_n)\cup V(H_n)$.
\item The edge set of $J_n$ is the union of the edge sets of $G_n$ and $H_n$ and one single extra edge in between
the vertices of $G_n$ and $H_n$. 
\end{itemize}
\noindent
Then, $\{J_n\}^\infty_{n=1}$ is still a hyperfinite sequence, since the expanders entails only a very small perturbation. 
So, all the reasonable graph properties are testable on the family $\{J_n\}^\infty_{n=1}$ in the uniform case. However,
it is very easy to construct a $K$-weighted sequence $\{J_n,p_n\}^\infty_{n=1}$ (for some large enough $K$) such that
half of the weight is concentrated on the expander part, which makes property testing impossible in most of the cases.
The bottom line is, small perturbation does not change the picture in the uniform case, but it could mean huge difference
if we wish to test the family against {\it all} possible distributions.
\vi
\section{Learning the Absolute Truth about Very Large Graphs} \label{absolute}
We learn graphs with distributions, building Alternate Realities. 
Is there an Absolute Truth?- we asked at the end of Section \ref{property}. Of course!- we would answer immediately, 
a graph $G$ is either a forest or not. For very large graphs, this is not so clear. In which possible, reasonable, practical, or even physical
sense a cycle of length $10^{10^{1000}}$ is {\it not} a forest? ``Property Testing With An Unknown Distribution'' offers a possible definition.
\begin{definition} A graph $G$ is $\epsilon$-close to a Property $\cP$ in the absolute sense, if for {\bf every vertex distribution} $p_G$,
$(G,p_G)$ is $\epsilon$-close to $\cP$.
\end{definition}
\noindent
Observe that this notion is not trivial. For any $\epsilon>0$ there exists some $N_\eps>0$ such that a cycle of length $n$ is $\epsilon$-close
to being a forest if $n\geq N_\eps$. Can we learn that a graph $G$ is $\eps$-close to some property in the absolute sense. Obviously, we cannot sample
against all possible distributions. We need to use some other way of learning: Learning by Observation.
Let $H\in \cG^d$ be any connected graph. Then, $Q_G(H)=YES$ if there exists at least one vertex $x\in V(G)$ such that $x$ is a vertex of an induced
subgraph isomorphic to $H$ and $Q_G(H)=NO$ if there is no such vertex. So by observation, we can collect qualitative data instead of quantitative data. \vi
Now, we introduce a strengthening of the notion of hyperfiniteness (a very similar but not equivalent
notion was introduced in our paper \cite{Elekqual}).
\begin{definition}[Uniform Hyperfiniteness]
A class of graphs $\cG$ of vertex degree bound $d$ is called {\bf uniformly hyperfinite} if for any $\epsilon>0$ there exists $K_\eps>0$, $L_\eps>0$
and for all $G\in \cG$ we have subsets $\{Y_i\}^{L_\eps}_{i=1}\subset V(G)$ such that
\begin{itemize}
\item For any $1\leq i \leq L_\eps$, $|V(Y_i)|<\eps |V(G)|$. 
\item For any $1\leq i \leq L_\eps$, if we delete $Y_i$ and all the incindent edges, the remaining graph has components at most $K_\eps$.
\item For all $x\in V(G)$,
$$\frac{|\{i\,\mid\, x\in Y_i\}|}{|L_\eps|}<\eps\,.$$
\end{itemize}
\end{definition}
\noindent
So, uniform hyperfiniteness is also sensitive to small perturbations.
The class of $D$-doubling graphs is uniformly hyperfinite \cite{Elekqual}. E.g. the family of all cycles is uniformly hyperfinite. 
Now, we give the precise definition of Observing a Property in a graph class $\cG$. This is the closest thing to ``Learning the Absolute Truth''.
We say that a {\bf graph property $\cP\subset\cG^d$ is observable } in the class $\cG$ if the following machinery works.
An Observing Oracle of depth $s$ takes a graph $G\in \cG$ and collects the truth values $Q_G(H)$ for all connected graphs $H$ such that $V(H)\leq s$.
So, we fix an $\epsilon>0$ and choose some $s_\epsilon>0$. Then an Observing Oracle of depth $s_\epsilon$ collects the truth values as above and
ACCEPT or REJECT the graph $G$ in such a way that:
\begin{itemize}
\item The Observing Oracle must ACCEPT $G$ if $G\in \cP$.
\item The Observing Oracle must REJECT $G$ if $G$ is not $\epsilon$-close to $\cP$ in the absolute sense.
\end{itemize}
\noindent
Note that observability is completely deterministic, it does not involve randomness at all. 
\begin{theorem}\cite{EK}
For a uniformly hyperfinite graph class any reasonable graph property is observable.
\end{theorem}

\vi
\begin{center} \textbf{\large The Continuous Part} \end{center}
The following sections are about the, hopefully interesting, connections in between the combinatorial ideas
of the previous sections and the theory of dynamical systems. \vi
\section{Boundary action of the free product group. Example 1. Revisited}\label{sec11}
Very large finite graphs are infinite in any practial sense, but they cannot be viewed as infinite graph, at least not in an easy way.
Let us consider an infinite, connected graph of $G$ vertex degree bound $10$. Try to pick a vertex $x$ of 
$G$ 
randomly uniformly. This is meaningless unfortunately, since if the vertices have the same measure and the
total measure is finite, then by $\sigma$-additivity, all vertices must have zero measure and the total
measure of the vertices must be zero as well. The following classical example from ergodic theory is
intimately related to our Example 1.
\begin{example} \label{eq2}
Let us consider the free product $\F$ of three copies of the cyclic group $\Z_2$. The elements
of $F$ can be identified with words consisting of letters $a$, $b$ and $c$ in such a way that
consecutive letters are always different. The multiplication operation is just the concatenation,
where consecutive $a$'s, $b$'s and $c$'s are cancelling each other. E.g. $abbac\cdot cacb=abbcb$,
the unit element is the empty word, and the inverse of the word $x_1x_2x_3\dots, x_n$ is 
$x_nx_{n-1}\dots x_1$. The Cayley graph $\Cay(\F)$ of $\F$ with respect to the generators $a,b,c$ is an infinite
$3$-regular tree, where each edge is labeled with the letters $a$, $b$ and $c$ in such a way that
adjacent edges have different labels.
\vi
The {\bf boundary} of $\F$ can be defined in the following way. Let $B_\F$ is the space
of all infinite sequences $x_1x_2x_2\dots$, where each $x_i$ is either $a$, $b$ or $c$ and all the consecutive
letter are different.
We equip $B_\F$ with the usual product topology (the topology of pointwise convergence) to obtain
the standard Cantor set. The compact set $B_\F$ is called the boundary of the group $\F$. 

\vi
Now we consider a natural action $\alpha:\F\actson B_\F$. That is, for each $\gamma\in F$ we
associate a homeomorphism $\alpha(\gamma)$ of $B_\F$ in such a way that $\alpha(\gamma\delta)=\alpha(\gamma)
\alpha(\delta)$. Let $w=y_1y_2y_3\dots y_n\in \F$ and $p=x_1x_2x_3\dots\in B_\F$. Then,
$\alpha(w)(p)=y_1y_2\dots y_nx_1x_2x_3\dots$ modulo the possible cancellations.
\vi
Let us consider the usual product (Lebesgue) measure $\mu$ on $B_F$. So, the measure $\mu(U_w)$ of the
open set $U_w$ of all the words starting with the word $w$ is $3^{-n}$, where $w=y_1y_2y_3\dots y_n$.
Then, $\alpha$ does not preserve the measure, but {\bf preserves its measure class}, that is, if for 
a measurable set $Z\subset B_\F$, $\mu(F)=0$, then for any $\gamma\in \F$, $\alpha(\gamma)(Z)=0$ as well.
These actions are called {\bf nonsingular actions}.
\end{example}
\section{The Radon-Nikodym Derivative} \label{secradon}
Let us briefly recall the notion of the Radon-Nikodym derivative.
Let $\Gamma$ be a countable group and $\beta:\Gamma\actson (X,\nu)$ is an action on a probability measure space $(X,\nu)$ preserving
the measure class of $\nu$. Here we can assume that $X$ is a standard Borel space (i.e. the interval or the
Cantor set) and the action maps Borel sets to Borel sets. Then, we have the following theorem.
\begin{theorem}[Radon-Nikodym]
Let $\beta:\Gamma\actson (X,\nu)$ as above.
Then, there exists a $\Gamma$-invariant subset $Y\subset X$ and a Borel function (the Radon-Nikodym Cocycle)
$R:\Gamma\times Y\to \R^+$ such that
\begin{itemize}
\item $\nu(Y)=1$.
\item For any $y\in Y$, $R(\gamma\delta,x)=R(\gamma,\beta(\delta)(x)) R(\delta, x)$.
\item For any Borel subset $A\subset Y$ and $\gamma\in\Gamma$
$$\nu(\gamma(A))=\int_A R(\gamma,x) d\nu(x)$$
\end{itemize}
\end{theorem}
\noindent
Note that up to zero measure perturbation the Radon-Nikodym Cocycle is unique.
\section{Limits of $K$-weighted graphs}\label{sec13}
Let us consider a finitely generated group $\Gamma$, with symmetric
generating system $\Sigma=\{\sigma_1, \sigma_2,\dots,\sigma_d\}$
and a nonsingular action $\alpha:\Gamma\actson (X,\mu)$ preserving the measure class
of $\mu$. We can assume that $X=X_0$ (as in the Radon-Nikodym Theorem).
Also, let us suppose that there exists $K>1$ such that
for all $x\in X$ and $\sigma_i\in\Sigma$
\begin{equation} \frac{1}{K}\leq R(\sigma_i,x)\leq K\,. \label{Keq} \end{equation}
\noindent
Then, for each $x\in X$, we can construct a countable, rooted, directed,
edge-labeled graph (the orbit graph) $G^\alpha_x$ of vertex degree bound $d$ in the
following way.
\begin{itemize}
\item $V(G^\alpha_x)$ is just the orbit of $x$, that is, $y\in V(G^\alpha_x)$ if there exists
$g\in\Gamma$ such that $\alpha(g)(x)=y$.
\item For each $\sigma_i\in\Sigma$ and $x\in X$, we have a directed edge
$\overrightarrow{(x,\beta(\sigma_i)(x))}$ labeled by $R(\sigma_i,x)$, where $R$ is the Radon-Nikodym Cocycle
(we do not consider loops or multiple edges).
\item The root of  $G^\alpha_x$ is $x$.
\end{itemize}
\noindent
Note that if we multiply the labels on the edges of a directed cycle we always get $1$. Also, if $\mu$ is actually
invariant under the action, then all the labels equal to $1$.
\vi
Now, let us consider again the example in Section \ref{sec11}.
The action $\alpha:\F\actson B_\F$ preserves the measure calls of $\mu$. It is
known that the support of the Radon-Nikodym Cocycle can be chosen in such a way that
each orbit graph $G^\alpha_x$ is a $3$-regular tree and from each vertex there is one outgoing
edge with label $2$ and two outgoing edges with label $\frac{1}{2}$.
\vi
In general, we can consider the set $C\cG_{d,K}$ of all rooted, connected graphs $G$ of vertex degree bound $d$,
equipped with an edge-labeling $l:\overrightarrow{E}(G)\to\R^+$ such that
\begin{itemize}
\item For every directed edge $\overrightarrow{e}$, $\frac{1}{K}\leq l(\overrightarrow{e})\leq K$.
\item The product of the labels on a directed cycle is always $1$.
\end{itemize}
\noindent
Now, let $(G,p_G)$ be a $K$-weighted graph of vertex degree bound $d$.
Then, we can construct a canonical probability measure $\mu_{p_G}$ on $C\cG_{d,K}$ in the following way.
For each $v\in V(G)$, we consider the edge-labeled rooted graph $(G,v,l_{p_G})$, where
$l(\overrightarrow{(a,b)}=\frac{p_G(b)}{p_G(a)}\,$ holds for adjacent vertices $a,b$.
We take $(G,v,l_{p_G})\in \cgdk$ with weight $p_G(v)$ to obtain the probability measure $\mu_{p_G}$.
Note that it is possible that $\mu_{p_G}$ is concentrated on less than $V(G)$ elements of $\cgdk$.
So, we can give the precise definition of the Weighted Benjamini-Schramm Convergence.
\begin{definition} \label{wbsc}
A sequence of $K$-weighted graphs $\{G_n,p_n\}^\infty_{n=1}$ with degree bound $d$ is
convergent in the sense of Benjamini and Schramm, if the sequence $\{\mu_{p_n}\}^\infty_{n=1}$ is convergent
in the weak topology of probability measures on $\cgdk$.
\end{definition}
\noindent
We can define the canonical probability measure $\mu_\alpha$ for a nonsingular action $\alpha:\Gamma\actson (X,\mu)$
as well.
Consider the Borel map $\Omega:X\to \cgdk$ defined by $\Omega(x)=G^\alpha_x$ and let $\mu_\alpha=(\Omega)_*(\mu)$, the push-forward
of the probability measure $\mu$.
\begin{definition} \label{limitw}
Let  $\{G_n,p_n\}^\infty_{n=1}$ be a convergent sequence of $K$-weighted graphs of vertex degree bound $d$.
The action $\alpha:\Gamma\actson (X,\mu)$ is the {\bf limit} of the sequence $\{G_n,p_n\}^\infty_{n=1}$, if
 $\{\mu_{p_n}\}^\infty_{n=1}$ converges to $\mu_\alpha$ in the weak topology.
\end{definition}
\noindent
Now, let us go back to our Example 1. Let $\{(T_n,p_n)\}^\infty_{n=1}$ be the sequence
of binary trees, where the probability measure is given by the critical value $\log(2)$.
Then, the sequence  $\{(T_n,p_n)\}^\infty_{n=1}$ is convergent and its limit is the action $\alpha:\F\actson B_\F$ of Example 2.
\vi
It is not hard to see that any convergent sequence of $K$-weighted graphs admits a limit action \cite{EK}. For the converse
we have the weighted version of the Aldous-Lyons ``Soficity'' Conjecture \cite{AL}.
\begin{conjecture}
If $\alpha:\Gamma\actson (X,\mu)$ is a nonsingular action of a finitely generated group with symmetric
generating system $\Sigma$ such that for any $\sigma\in\Sigma:$, we have that $\frac{1}{K}\leq R(\sigma,x)\leq K$.
Then, there exists a sequence of $K$-weighted graphs $\{G_n,p_n\}^\infty_{n=1}$ such that $\alpha$
is the limit of $\{G_n,p_n\}^\infty_{n=1}$ .
\end{conjecture}
\vi
\section{Hyperfinite Graphs vs. Hyperfinite Actions}
First let us recall the classical notion of $\mu$-hyperfiniteness.
Again, let $(\Gamma,\Sigma)$ be a finitely generated group
with a symmetric generating set and let  $\alpha:\Gamma\actson (X,\mu)$ be a nonsingular action.
The action is called {\bf $\mu$-hyperfinite} if 
for any $\eps>0$ there exists $Y\subset X$ such that
\begin{itemize}
\item $\mu(Y)<\epsilon$.
\item If we delete the points of $Y$ and all edges incident to $Y$ in the orbit graphs $G^\alpha_x$, all the
remaining components have size at most $K_\eps$.
\end{itemize}
Note that $\mu$-hyperfiniteness does not depend on the choice of $\Sigma$. All nonsingular actions
of amenable groups are $\mu$-hyperfinite. None of the measure preserving essentially free actions (the fixed point
set of any nontrivial element has measure $0$) of nonamenable
groups are hyperfinite and for all groups $\Gamma$ there exist $\mu$-hyperfinite essentially free actions that are 
$\mu$-hyperfinite.  The following theorem is the nonsingular analogue of Schramm's Theorem \cite{Schramm}.
\begin{theorem}
Let $\{G_n,p_n\}^\infty_{n=1}$ be a convergent sequence of $K$-weighted graphs and let $\alpha:\Gamma\actson (X,\mu)$ be
a limit action of $\{G_n,p_n\}^\infty_{n=1}$. Then, $\{G_n,p_n\}^\infty_{n=1}$ is weighted hyperfinite if and only if
$\alpha$ is $\mu$-hyperfinite. 
\end{theorem}
\noindent
Also, any $\mu$-hyperfinite action $\alpha:\Gamma\actson (X, \mu)$ satisfying \eqref{Keq} is a limit action of
some convergent sequence of $K$-weighted graphs. 
E.g. the action in Example 2. $\alpha:\F\actson (B_\F,\mu)$ is $\mu$-hyperfinite and of course
the sequence in Example 1.  $\{T_n,p_n\}^\infty_{n=1}$ that tends to $\alpha$, is weighted hyperfinite as well.
\section{Continuous, Free, $\mu$-hyperfinite Actions  via $K$-weighted graphs}
The following construction is a small modification of our Example 1.
Let $(\Gamma,\Sigma)$ be a finitely generated nonamenable group with a symmetric generating system.
Let $B_n$ denote the ball of radius $n$ around the unit in the left Cayley graph of $\Gamma$ with respect
to $\Sigma$ and let $S_n$ be
the $n$-sphere, that is, the boundary of $B_n$.  Then, the growth entropy 
$$\beta=\lim_{n\to\infty} \frac{\log(|B_n|)}{n}=\lim_{n\to\infty} \frac{\log(|S_n|)}{n}$$
\noindent
exists and greater than zero (note that the growth is positive for some amenable groups as well). 
Suppose that there exists some $C>0$ such that
\begin{equation}\label{coorn}
\frac{1}{C} e^{\beta n}\leq |B_n| \leq C e^{\beta n}
\end{equation}
holds for all $n\geq 1$. E.g. the so-called hyperbolic groups  satisfy \eqref{coorn} by a result of Coornaert \cite{Coornaert}. 
\vi
Then, we can repeat the construction in Example 1. basically word by word.
Let $w_n: B_n\to\R^+$ be defined by $w_n(x)=e^{-\beta k}$, where
$\dist_{B_n} (e,x)=k$ and let $p_n(x)=\frac{w_n(x)}{\sum_{y\in B_n} w_n(y)}\,.$
Then, as in Example 1, we have that
$\lim_{n\to\infty} p_n(S_n)=0$, so for any $r>0$ if $n$ is large enough, the ball around a random vertex look like the
ball $B_r$.  If \eqref{coorn} does not hold we slightly modify our construction.
Let $w_n(x)=\frac{1}{|S_k|}$, if $d_{B_n}(e,x)=k$ and
if $d_{B_n}(e,x)=k$ then let $p_n(x):=\frac{w_n(x)}{\sum_{y\in B_n} w_n(y)}\,.$
Again, we have that $\lim_{n\to\infty} p_n(S_n)=0$.
We already mentioned Sako's result \cite{Sako} on Property $A$ in Section \ref{sec9}. Now, we can apply this theorem again,
to show that if $\Gamma$ is a so-called exact group (say, hyperbolic or amenable) then the sequence $\{(B_n,p_n)\}^\infty_{n=1}$ is weighted hyperfinite.
One should note, that subgroups, extensions, free products and directed unions of exact groups are exact as well, so
groups have a tendency being exact. 
The first nonexact group was constructed only in 2003 by Gromov \cite{Gromov}.
The following theorem follows relatively easy from our construction above \cite{EK}.
\begin{theorem}
Let $\Gamma$ be a finitely generated exact group. Then, there exists a free, minimal continuous action
$\alpha:\Gamma\actson \cC$ on the Cantor set preserving the measure class of $\mu$ such that
\begin{itemize}
\item The action is amenable and purely infinite (this has been done by R{\o}rdam and Sierakowski \cite{Ror}).
\item All the Radon-Nikodym derivatives are continuous, taking values from the integer powers of $2$.
\end{itemize} \end{theorem} \vi

\section{A Very Small Remark on the Furstenberg Entropy}
If $(G,p_G)$ is a $K$-weighted graph, then the most important invariant of $p_G$ is its entropy 
$$H(p_G)=\sum_{v\in V(G)} -\log(p_G(v)) p_G(v)\,.$$
\noindent
Note that for our very large $K$-weighted graphs the entropy, in general, very big.
However, we can consider the {\bf edge-entropy} $H_{edge}(G,p_G)$ of $(G,p_G)$ defined as
$$\sum_{v\in V(G)} \sum_{e_out(x)} -\log (r(e)) p_G(x)\,,$$
\noindent
where $e_{out}$ is the set of edges outgoing from $x$, and $r(\overrightarrow{(x,y)})=\frac{p(y)}{p(x)}$ is
the Discrete Radon-Nikodym Derivative in the direction of $y$. 
Clearly, $H_{edge}(G,p_G)< d \log(K)$. Also, by the definition of the weighted Benjamini-Schramm convergence, if
$\{(G_n,p_n)\}^\infty_{n=1}$ is convergent, then $\{H_{edge}(G_n,p_n)\}^\infty_{n=1}$ is convergent. Arguably, $H_{edge}$ is one of the most natural 
testable parameters for general $K$-weighted groups.
Let us suppose that $\alpha:\Gamma\actson (X,\mu)$ is an essentially free, nonsingular action of a finitely generated group 
$\Gamma$ with symmetric generating system $\Sigma$
and $\alpha$ is the limit of the convergent sequence $\{(G_n,p_n)\}^\infty_{n=1}$ of $d$-regular graphs. 
Then, by definition
$$\lim_{n\to\infty} H_{edge}(G_n,p_n)= \sum_{\sigma\in\Sigma} \int_{X} -\log ( R(\sigma,x)) d\mu(x)\,.$$
The last quantity (or rather $\frac{1}{d}$-times this quantity) is called the Furstenberg-entropy of the action and it
has already been defined in $1963$ \cite{Furstenberg}.

\end{document}